\documentclass[twocolumn]{svjour3}

\usepackage{amsmath,amsfonts,amssymb}
\usepackage{setspace}
\usepackage{tocloft}
\usepackage[flushleft]{threeparttable}
\usepackage{natbib}
\usepackage{graphicx}
\usepackage{lipsum}
\usepackage{float}
\usepackage{amsmath}
\usepackage{latexsym}
\usepackage{booktabs,etoolbox}
\usepackage[utf8x]{inputenc}
\usepackage{caption}

\newcommand{\as}{\prime\prime}
\newcommand{\D}{\Delta}

\def\be{\begin{equation}}
\def\ee{\end{equation}}
\def\fr{\frac}

\newcommand{\apj}{\it Astrophys.~J.}
\newcommand{\apjl}{\it Astrophys.~J.~Lett.}
\newcommand{\mnras}{\it Mon.~Not.~R.~Astron.~Soc.}

\begin{document}

\title{Wide-field Ultraviolet Imager for Astronomical Transient Studies}

\author{Joice Mathew \and S. Ambily \and Ajin Prakash \and Mayuresh Sarpotdar \and Nirmal K. \and A.G. Sreejith  \and  Margarita Safonova \and Jayant Murthy\and Noah Brosch}

\institute{Joice Mathew\and S. Ambily \and Ajin Prakash \and Mayuresh Sarpotdar   \and  Nirmal K. \and Jayant Murthy\at
Indian Institute of Astrophysics, Koramangala 2nd block, Bangalore, 560034, India \\
              \email{joice@iiap.res.in}        \\
           \and
           A.G. Sreejith \at Space Research Institute, Austrian Academy of Sciences, Schmiedlstrasse 6, Graz, Austria \and  Margarita Safonova \at M.~P.~Birla Institute of Fundamental Research, Bangalore, India \and 
Noah Brosch\at
The Wise Observatory and the Dept. Of Physics and Astronomy, Tel Aviv University, Tel Aviv 69978, Israel}

\date{Received: date / Accepted: date}

\maketitle

\begin{abstract}

Though the ultraviolet (UV) domain plays a vital role in the studies of astronomical transient events, the UV time-domain sky remains largely unexplored. We have designed a wide-field UV imager that can be flown on a range of available platforms, such as high-altitude balloons, CubeSats, and larger space missions. The major scientific goals are the variability of astronomical sources, detection of transients such as supernovae, novae, tidal disruption events, and characterizing AGN variability. The instrument has a 80 mm aperture with a circular field of view of 10.8 degrees, an angular resolution of $\sim$22 arcsec, and a $240 –- 390$ nm spectral observation window. The detector for the instrument is a Microchannel Plate (MCP)-based image intensifier with both photon counting and integration capabilities. An FPGA-based detector readout mechanism and real time data processing have been implemented. The imager is designed in such a way that its lightweight and compact nature are well fitted for the CubeSat dimensions. Here we present various design and developmental aspects of this UV wide-field transient explorer.

\end{abstract}


\section{Introduction}
 
We have a limited view on  how the UV sky looks because of the lack of current UV missions. This spectral region has a special significance in astronomy as it contains a wealth of information on the physical properties of gas, stars and galaxies \citep{2014AdSpR..53..982C}. Indeed, this region is critically important for characterizing the emission from young and/or hot stars, as well as from hard non-thermal sources. Also, the UV spectral range is blessed with a very dark sky background, which permits achieving a good signal to noise (S/N) ratio on faint sources with relatively small optics and with a sensitive low noise detector.

The time-domain astronomy is rarely explored in the UV even though it can have potential scientific impact on the study of transient events \citep{Broschsmall,2014AJ....147...79S}. In most electromagnetic bands, the static sky has been studied in depth. Studies related to cosmic explosions require wide-field time-domain imaging surveys \citep{2014AJ....147...79S}. Technological advances in detectors, computing power, data storage capability, and astronomical data analysis enable us to efficiently monitor large areas of sky searching for transient events. Examples include the discovery of rare or unusual transients \citep{2012ApJ...753...77C}. Astrophysical transient events include bursts, flashes, flares, SNe, gamma-ray bursts (GRB), etc. The rate of discoveries of transient events is constantly
increasing with larger telescopes and newer techniques coming on line.
Both SWIFT and GALEX have been yielding impressive results in detecting transients (e.g. \citep{Botticella2010,Rabinak2011}. The discovery rate of transients in the UV could increase by several orders of magnitude if a dedicated wide-field UV instrument would be launched \citep{Gezari2013}.

Transient events have been so far studied by the Palomar Transient Factory (PTF) in the optical domain, by LOFAR and SKA in radio, and by SWIFT, Fermi Gamma-ray Space Telescope and LOFT in the high-energy region. But no UV wide-field imager dedicated for transient studies has yet been flown, even though it could address major scientific questions in astronomy. No wide-field synoptic imaging in the UV has been done to detect serendipitous sources, with the exception of GALEX \citep{2005ApJ...619L...1M} where six supernovae were detected in the NUV channel during a very limited time-domain program \citep{2011A&A...527A..15W, 2016ApJ...820...57G}. The only UV instruments now operating in space are the HST, UVIT (Ultraviolet Imaging Telescope) and the auxiliary telescopes on SWIFT and XMM-Newton, though none of them can perform the required wide-field imaging. One main advantage of the UV monitoring is an early detection of a UV flash, which can interpreted as the shock break-out of a supernovae explosion \citep{2010ApJ...725..904N}. Synoptic monitoring in the UV can also enhance understanding of the flaring mechanisms of dwarf stars, and of tidal disruption events (TDEs) \citep{Broschsmall}. 

We have developed a wide-field UV imager (WiFI) capable of transient survey in the NUV domain. The major scientific goals of this instrument are the studies of massive star explosions, tidal disruption events (TDEs), variability studies of active galactic nuclei 
(AGNs), various classes of variable stars, and electromagnetic counterparts to gravitational-wave (GW) sources \citep{2014AJ....147...79S}.

\section{Instrument Overview}

The instrument is a UV imaging refractive telescope with 10.8 degrees field of view (FOV). The detector is an MCP-based image intensifier which can work both in photon-counting and integrating mode, and has a good sensitivity in the NUV. The detector has a spectral response over the 200--800 nm range; therefore we added a UV bandpass filter (Hoya U-235C) at a distance of 15 mm before the detector focal plane, to restrict the bandpass to 240–-390 nm. This UV filter has a peak transmission of $\geq 88\%$ at 320 nm, and has an excellent cut-off in the visible band.

\begin{table}[h]
\caption{Instrument Details}
\begin{center}
\small
\begin{tabular}{ll}
\hline 
 Instrument & UV wide-field imager \\
 Telescope type & Lens system  \\
 Field of view & $10.8^{\circ}$ (circular)\\
 Aperture diameter  &  80 mm  \\
 Focal length & 209.1 mm  \\
 Operating bandwidth & $240-390$ nm\\
 Detector & Micro channel plate (MCP)  \\
 Pore size & $10\,\mu$m \\
 Spatial resolution & $\sim 22^{\prime\prime} $\\
 Time resolution & 30 msec \\
 Sensitivity & 18 AB (100 seconds exposure)\\
Weight & $ <4$ kg \\
 Dimension (L $\times$ W$\times$ H)& $280 \times 160\times90$ mm \\
 & (unfolded design) \\
Power & $ < 5$  W \\
\hline
\end{tabular}
\label{table:instrument details}
\end{center}
\end{table}
The instrument is designed in such a way that it can be flown on a range of platforms, such as high-altitude balloons, CubeSats, or nanosatellites. The optical design could be slightly modified so that the complete instrument can fit in a 6U CubeSat (folded design, Sec.~6.2, Fig.~13). During its operation, the instrument will always look in anti-Sun direction. The imager can cover a large portion of the sky and acquire images for long times; these images will be stored and scanned to look for any brightness variation. 
The specifications of the instrument are given in Table~\ref{table:instrument details}.

\section{Optical Design}

\begin{figure*}[h!]
\begin{center}
\includegraphics[scale=0.4]{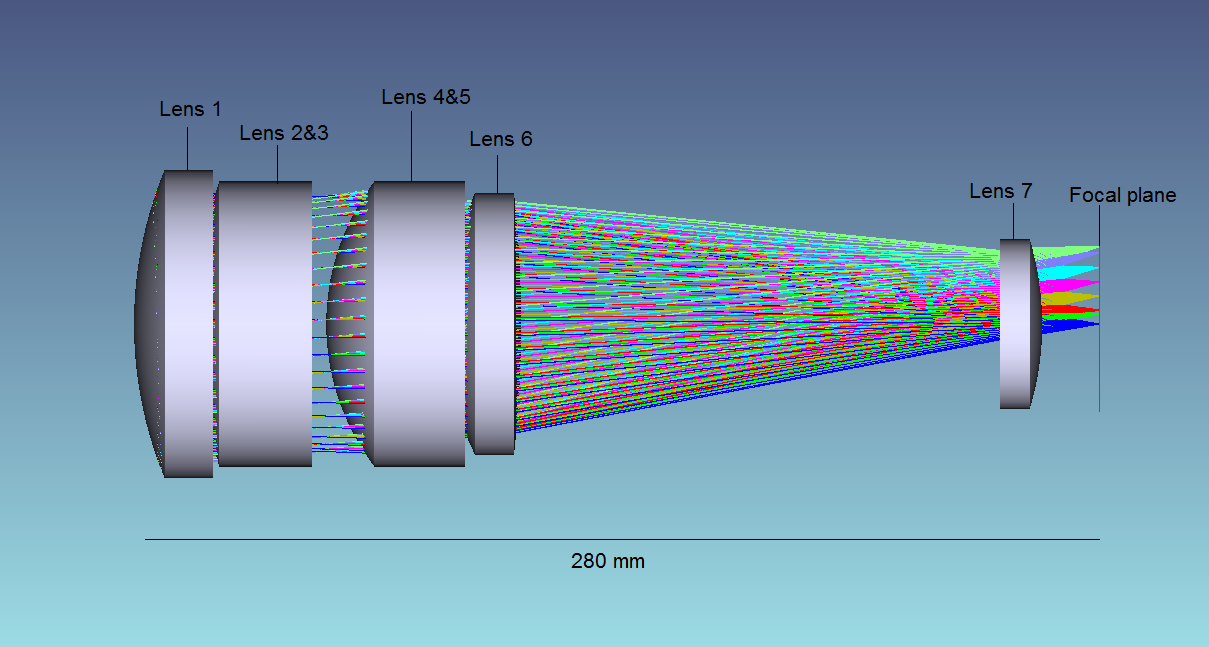}
\end{center}
\caption{Optical layout of wide-field UV imager}
\label{fig:Optical Layout}
\end{figure*}

The optical system consists of seven lenses made of fused silica and calcium fluoride ($CaF_2$). For ease of manufacture, we have considered all lens surfaces to be spherical and all surfaces  to be anti-reflection coated with $MgF_2$. The second and third lenses are glued and form a doublet, and the same for the fourth and fifth lenses (also a glued doublet). The optical design was carried out using Zemax, and the optical layout is shown in Fig.~\ref{fig:Optical Layout}.

\begin{figure}[h!]
\begin{center}
\includegraphics[scale=0.25]{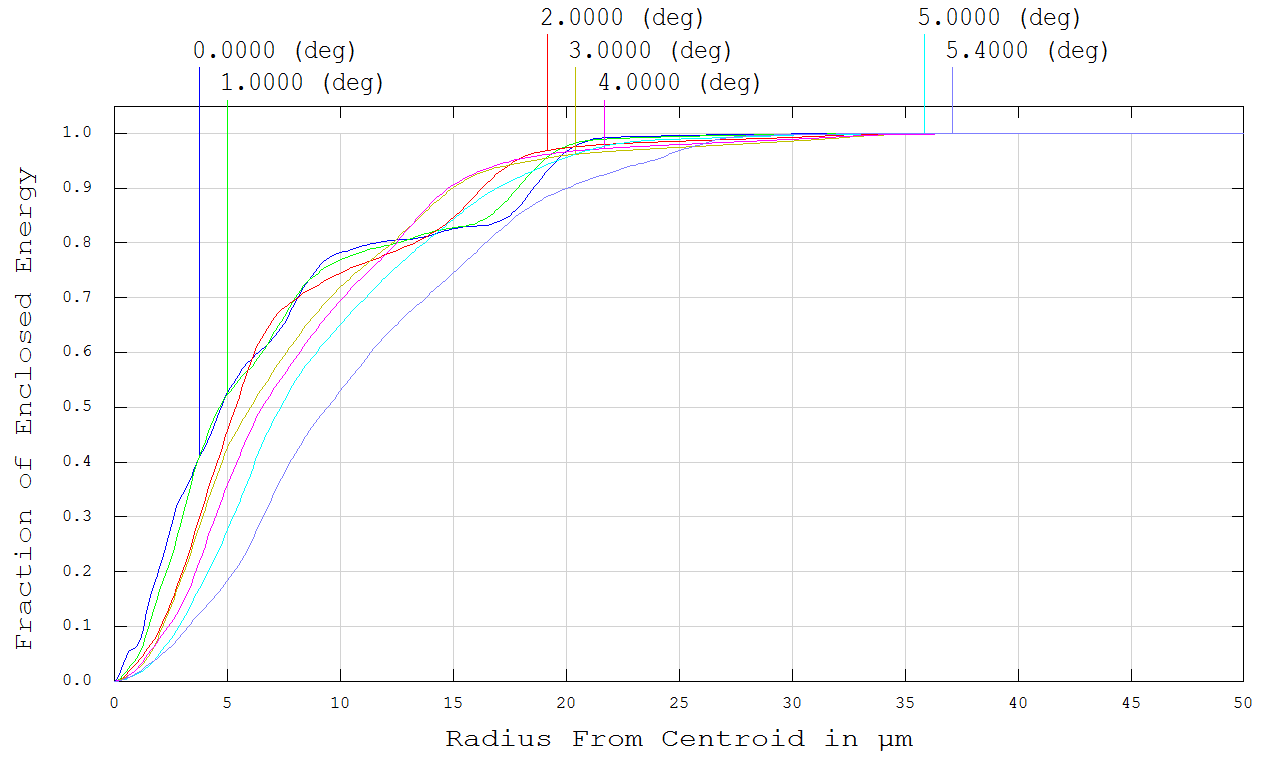}
\end{center}
\caption{Diffraction-encircled energy for on-axis and off-axis fields. Colours indicate different fields, with offset from the optical axis shown in the plot.}
\label{fig:Encircled Energy}
\end{figure}

\begin{figure}[h!]
\begin{center}
\includegraphics[scale=0.3]{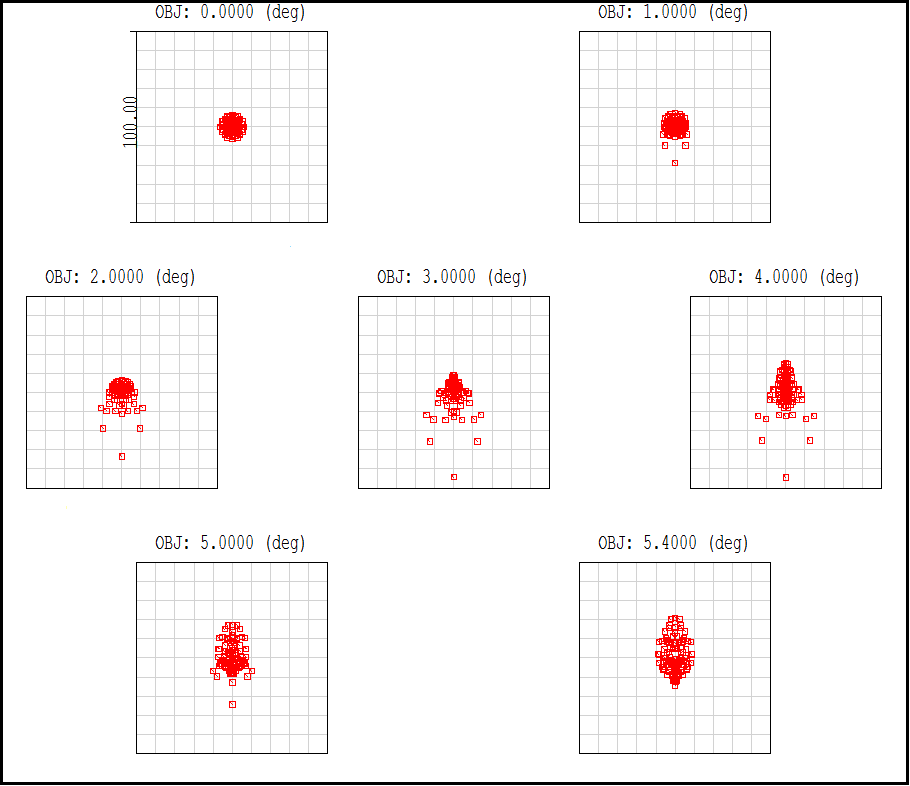}
\end{center}
\caption{Spot diagram for on-axis and off-axis fields. Each small square in the spot diagram is of 10 $\mu$m x 10 $\mu$m size.}
\label{fig:spot}
\end{figure}

Major constraints in the designs are the availability of UV transparent glasses and the spherical nature of the optical surfaces. The imaging performance criteria were to have at least 70\% of the encircled energy (Fig.~\ref{fig:Encircled Energy}) should fall within three pores of MCP after considering the manufacturing and alignment tolerances. We have optimized the radii of curvature, thickness of lenses and spacing between the lenses for the desired optical performance. The aperture of the system is 80 mm with a focal length of 209.1 mm. For the on-axis field, the RMS spot radius is 5.2 $\mu m$, and for the most distant field at  5.4$^\circ$ from the center, the RMS spot radius is 11.9 $\mu m$ (Fig.~\ref{fig:spot}).

To evaluate the performance of the opto-mechanical
system, we carried out a tolerance analysis. We have
combined the sensitivity analysis and Monte Carlo simulations, using Zemax, to investigate the degradation of the
encircled energy. The tolerances for manufacturing and
alignment are shown in Table~\ref{table:tolerance values}. 

\begin{table}[h!]
\begin{center}
\caption{Tolerance allocation on manufacturing and alignment}
\small
\begin{tabular}{lllc}
\hline
Tolerance term & sub tolerance term &  tolerances\\
\hline
Manufacture & Radius of curvature &  0.1\% \\
  & Thickness ($\mu$m) & $\pm 100$\\
  & Decenter in X $\&$ Y ($\mu$m) &$\pm 50$\\
  & Tilt in X $\&$ Y ($^{\prime\prime}$)  & 60\\
  & Surface irregularity  &  $\Lambda/6$\\
\rule[-1ex]{0pt}{3.5ex} Alignment & Decenter in X $\&$ Y ($\mu$m) &  $\pm 50$\\
             & Tilt in X $\&$ Y ($^{\prime\prime}$) & 60\\
\hline
\end{tabular}
\label{table:tolerance values}
\end{center}
\end{table}

The back focal length of the
telescope is used as the compensator.
The results of this tolerance analysis showed that the system performance can
be achieved easily with moderate tolerances on the fabrication and assembly. As a result of the moderate tolerance requirements and the smaller number of parts, this would reduce significantly the cost of the system.

\begin{figure}[h]
\begin{center}
\includegraphics[scale=0.25]{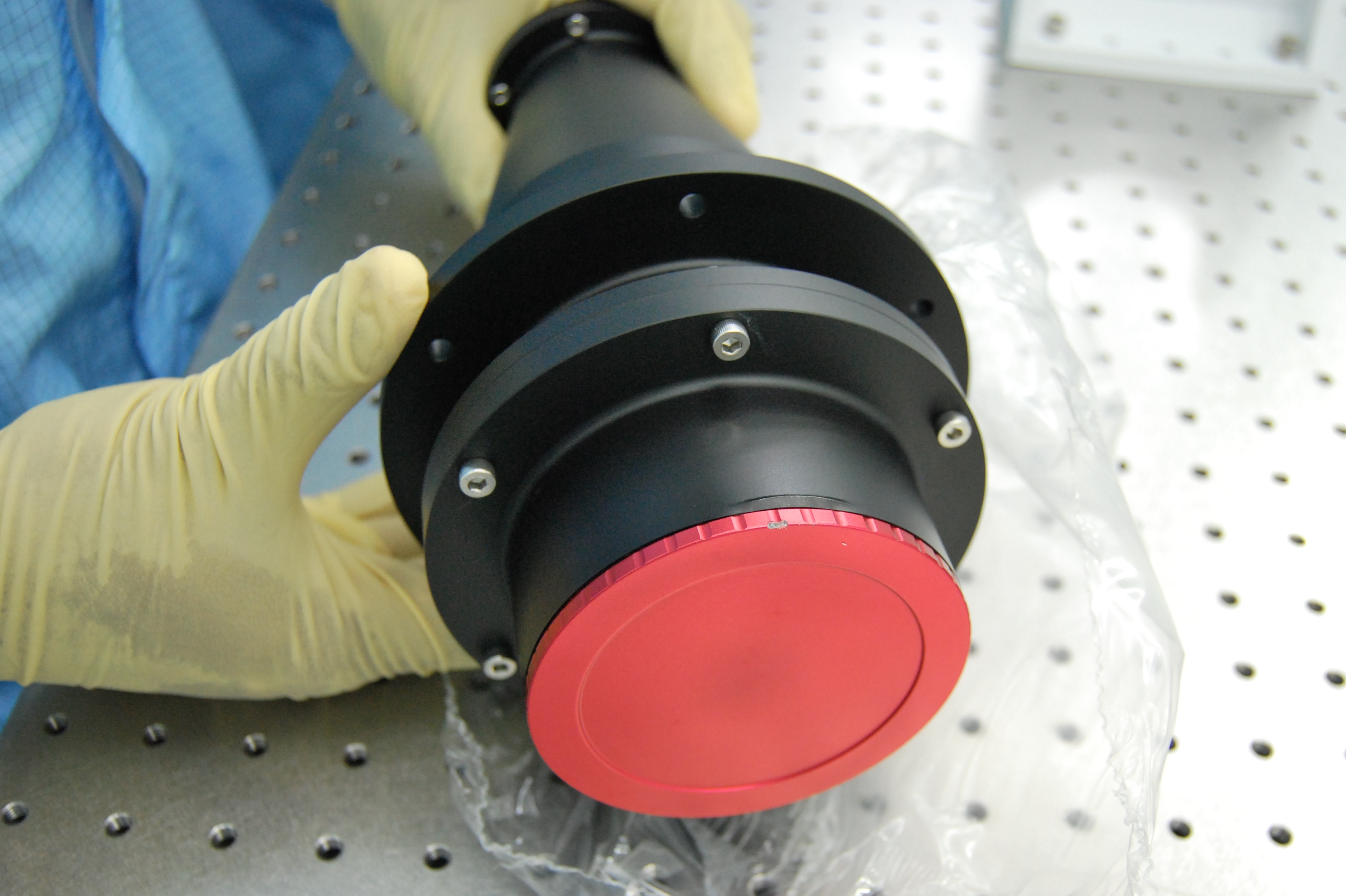}
\end{center}
\caption{Assembled lens system.}
\label{fig:Lens system}
\end{figure}

Ghost image analysis was also performed in Zemax using ghost image focus generator option. We found that no ghost image is formed at the focal plane, and the nearest ghost image focus is 60 mm before the detector plane. Since the lens is AR-coated (with 90\% transmission in NUV and  reflectivity less than 0.5\%  for each optical surface), the contribution of flux due to ghost rays is small, around 0.2\% of the actual image flux on the detector plane. 

The imager requires outer baffles to reduce the straylight, mainly from the Sun,the Moon, and the bright Earth limb. The internal walls of the optical tube have been black anodized to reduce the scattering. However to fly on a balloon, the baffles are not required since the observations will be done during the local night, and the imager will always be pointed away from the Moon. For a CubeSat platform, a deployable baffle \citep{CubeSat_baffle} will be required. The assembled lens system is shown in Fig.~\ref{fig:Lens system}

\section{Detector and Electronics}

It is possible to build position-sensitive detectors in the UV using micro channel plates (MCP), where photo-electrons are accelerated and multiplied within individual channels (pores)\citep{2009Ap&SS.320..247V}. With three-stack MCPs, gains of $10^6$ to $10^7$ can be achieved. The resulting electron cloud can be read out with an anode or a phosphor screen. In optical readout, the phosphor output is activated by the electron cloud emerging from the last MCP of the stack, with the phosphor optically-coupled to a regular, fast readout CMOS. Using fast analyzing electronics hardware, frames can be read at a high rate and centroiding can be done on-board, yielding sub-pixel accuracy. The resultant photon counting detector module has high spatial and temporal resolution. The operation of similar detectors have been proved in the past, for example the  detectors on UVIT mission on-board ASTROSAT \citep{Kumar}.

\begin{figure*}[h]
\begin{center}
\includegraphics[scale=0.37]{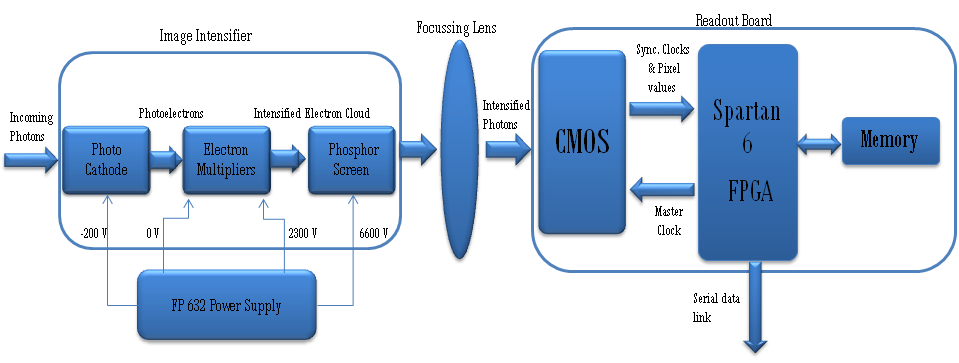}
\end{center}
\caption{MCP-based UV detector block diagram}
\label{fig:MCP-based UV detector block diagram}
\end{figure*}

We have developed a compact detector using off-the-shelf MCP (micro-channel plates), CMOS(complement-ary metal-oxide semiconductor) sensors and of-the-shelf optics \citep{ambily}. We are using the MCP340 assembly from Photek\footnote{\tt{http://www.photek.com/}}, which is a 40 mm diameter 3-stage MCP. It includes a quartz input window, S-20 photocathode, Z-stacked MCP and a P46 phosphor screen. The MCP requires a HVPS to operate in a single-photon mode, for which we have used the FP632, a micro-HVPS from Photek weighing only 300 gm, which makes it attractive for use on light-weight missions. The MCP voltage levels are $-200 V, 0 V, 2300 V$ and $6600 V$ for the cathode, MCP-In, MCP-Out and phosphor screen, respectively. It is possible to adjust the gain of the MCP by changing the voltage at the MCP output, which automatically adjusts the screen voltage, thus maintaining a constant MCP-out-Anode voltage difference. The phosphor screen has a peak response at 530 nm, and the resulting photons are focused on the CMOS surface through a focussing lens. The UV detector block diagram is shown in Fig.~\ref{fig:MCP-based UV detector block diagram}.  

Similar commercial systems often use a fiber-optic coupling but these are more difficult to physically implement with high precision. We are now using the TVF-3.6X2812IR-BCDN-MD, a board-level varifocal le-ns from Senko with high precision zoom and focus adjustment, which was provided with the CMOS board.

For the optical readout of the MCP, we are using an off-the-shelf CMOS headboard (Soliton Technologies) which has OV9715, a 1 Mega Pixel Video Image Sensor from OmniVision Technologies, as the image sensor and some essential components for biasing and clocking the CMOS sensor. It is a 1/4 inch sensor that provides full-frame, sub-sampled or windowed 8-bit/10-bit images in RAW RGB format via the digital video port and with complete user control over image quality, formatting and output data transfer. The headboard reads the digital image values and sends them to the FPGA for further processing. It also incorporates a synchronous serial port, which may be used to modify the internal camera registers when needed. 
The pixel stream and the three sync signals for the pixel, line, and frame clocks are generated by the CMOS sensor and are connected directly to the FPGA inputs. The master input clock for the CMOS chip is generated by the on-chip crystal oscillator and PLL on the FPGA board. The oscillator frequency is fixed at 12 MHz, but the input clock frequency for the CMOS chip may be varied from 6 to 27 MHz using the programmable PLLs on the FPGA. 

\begin{figure}[h]
\begin{center}
\includegraphics[scale=0.4]{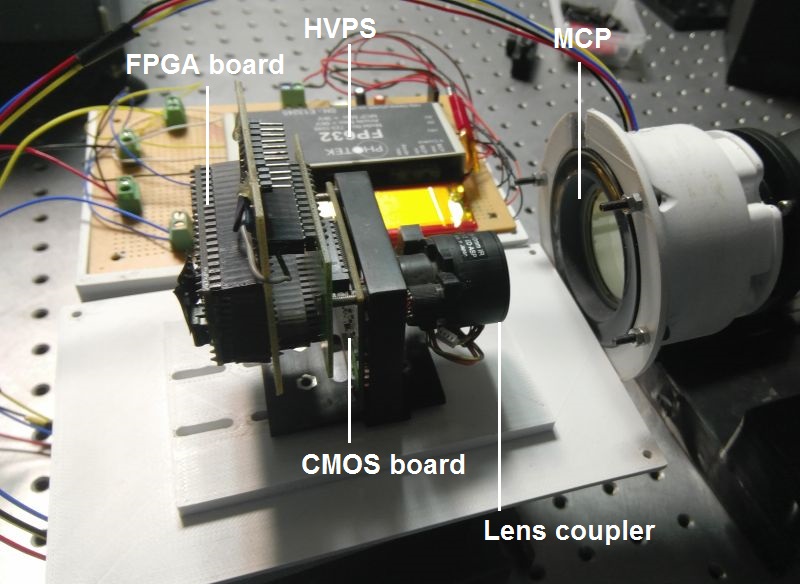}
\end{center}
\caption{MCP-based UV detector assembly}
\label{fig:MCP-based UV detector}
\end{figure}

For the design of the main processor, we have chosen an FPGA prototyping board, XuLA2-LX25 from XESS Corp. with a Spartan 6 series FPGA (XC6LX25) from Xilinx, Inc. The FPGA is designed and programmed in such a way that it can serve the multiple functions such as control and readout of the CMOS sensor, processing and computation of centroids, and storage and transmission of the output. The inherent advantages of FPGAs, such as highly customizable architecture, the ability to do parallel processing, faster lead times, and lower costs have helped us in making changes in the algorithms without major modification of the overall system architecture. The detector assembly is shown in Fig.~\ref{fig:MCP-based UV detector}.

The readout electronics supports two modes of operation viz. photon counting and frame transfer and can switch between these two modes without any changes in the hardware. In the two modes of operation specified, the readout of the CMOS sensor does not differ much. In the photon counting mode, the incoming photons are first converted to photoelectrons, and this electron cloud is amplified by a series of electron multipliers in the MCP. This cloud is further converted to photons at a peak wavelength 530 nm when it falls on a phosphor screen. The resulting amplified photon cloud is read out by the CMOS sensor, and the exact centroid of each photon event is calculated for better temporal and spatial resolution. The output in this mode will be a photon event list, which will have the $x$ and $y$ coordinates of the event on the detector and a time stamp.  

In continuous frame transfer mode, the working is more or less straightforward. The pixel values and the synchronization clocks are read out continuously by the FPGA chip and then stored in the SDRAM temporarily. Once a complete frame is read, the controller stops reading the CMOS chip. It then commences reading back the data from the RAM and writing it on the micro SD card present on-board and/or transmitting the data over the serial port. So far, we have implemented asynchronous RS232 communication which is simple but rather slow, although it is possible to use any of the other synchronous serial transmission at a later stage. Each 16-bit data packet consists of an 8-bit value for the pixel count from the sensor, and 4 bits for the frame and row ID each. From the received data packets it is possible to reconstruct the original image, using any of the image processing software such as, e.g. MATLAB.

\begin{figure}[h!]
\begin{center}
\includegraphics[scale=0.6]{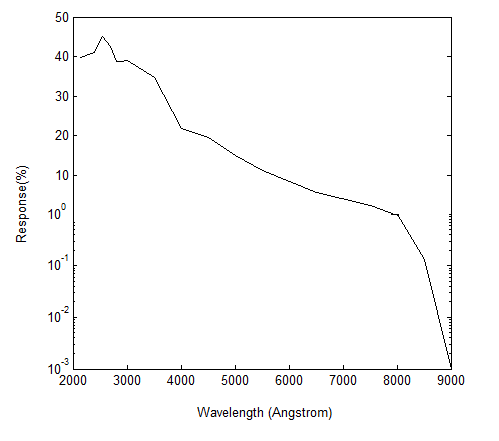}
\end{center}
\caption{Detector quantum efficiency plot with a logarithmic scale longward of 4000 \AA.}
\label{fig:qe}
\end{figure}

\begin{figure}[h!]
\begin{center}
\includegraphics[scale=0.6]{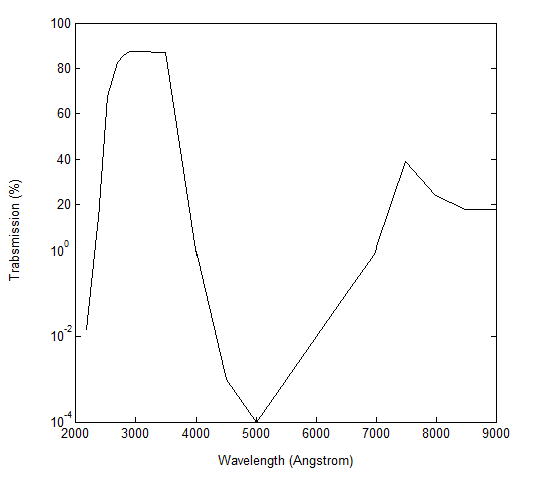}
\end{center}
\caption{Filter response plot with a logarithmic scale longward of 4000 \AA.}
\label{fig:Filter}
\end{figure}

The S20 photocathode has more than 30\% QE in the desired wavelength band and its also sensitive in visible band. To achieve the required pass band we have used a visible blocking bandpass filter. The detector and filter responses are shown in Fig.~\ref{fig:qe} and Fig.~\ref{fig:Filter} respectively, based on the measured data provided by the vendors.

For the definition of the imager's photometric system, we adopt the AB magnitude system widely used in UV astronomy \citep{OkeGunn},
\be
m_{\rm {WiFI}} = -2.5\log{\langle F_{\nu}\rangle} -48.6\,,
\ee
where $\langle F_{\nu}\rangle$ is averaged across the bandpass monochromatic flux (ergs/sec/cm$^2$/Hz) at central (mean) wavelength $\lambda_0$,
\be
\lambda_0= \fr{\int \lambda R_{\lambda} d\lambda}{\int R_{\lambda} d\lambda}\,,\quad \langle F_{\nu}\rangle = \fr{\int f_{\nu} R_{\nu} 
d\nu}{\int R_{\nu} d\nu} \,, 
\ee
and where
\be
\langle F_{\lambda}\rangle = \fr{\int f_{\lambda} R_{\lambda} d\lambda}{\int R_{\lambda} d\lambda}\,\,,f_{\lambda}=f_{\nu}\fr{c}{\lambda^2_{\rm p} \cdot 10^8}\,\,,R_{\nu}=R_{\lambda}\,.
\label{eq:2}
\ee

\begin{figure}[h]
\begin{center}
\includegraphics[scale=0.65]{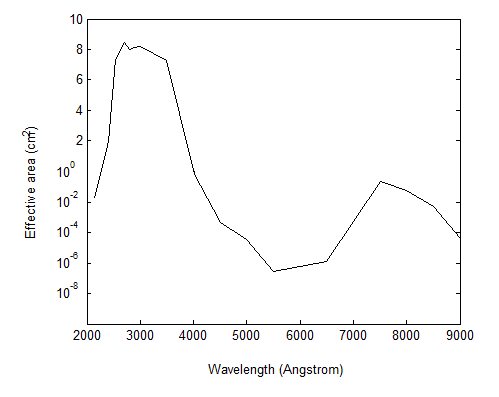}
\end{center}
\caption{Effective area curve. We have used a logarithmic scale longward of 4000 \AA\, to emphasize the red light rejection.}
\label{fig:effective_area}
\end{figure}

We express the total system response characterizing the instrument’s  efficiency in transmitting light in terms of effective area in cm$^2$, 
\be
A_{\rm eff}
= A_{\rm coll}\times T_{\rm l} \times T_{\rm f}(\lambda)\times {\rm QE}(\lambda)\,,
\ee
where $A_{\rm coll}$ is the effective geometrical collecting area, $T_{\rm f}(\lambda)$ and $T_{\rm l}$ -- filter and lens transmission in NUV, respectively, and QE$(\lambda)$ is the quantum efficiency of the detector. The effective area plot for the instrument is shown in Fig.~\ref{fig:effective_area}.

The instrument will be calibrated in a class 1000 clean environment, at the M.~G.~K. Menon Space Science Centre at the CREST campus of the Indian Institute of Astrophysics, Bangalore, India. This facility was used for the integration, characterization and calibration of the UVIT instrument \citep{Kumar,UVIT_2017}. To avoid any contamination, we have selected low outgassing materials and performed vacuum bake-out, ultrasonic cleaning and/or solvent cleaning of structural components. The complete instrument will be purged with ultra-clean nitrogen after the initial assembly till few days before the launch, to avoid contaminations.

\section{Mechanical Design}

We have carried out the mechanical design in such a way that in can handle all launch vibrations and shocks during launch with as a minimal weight as possible. 
\begin{figure*}[h!]
\begin{center}
\captionsetup{justification=centering}
\includegraphics[scale=0.40]{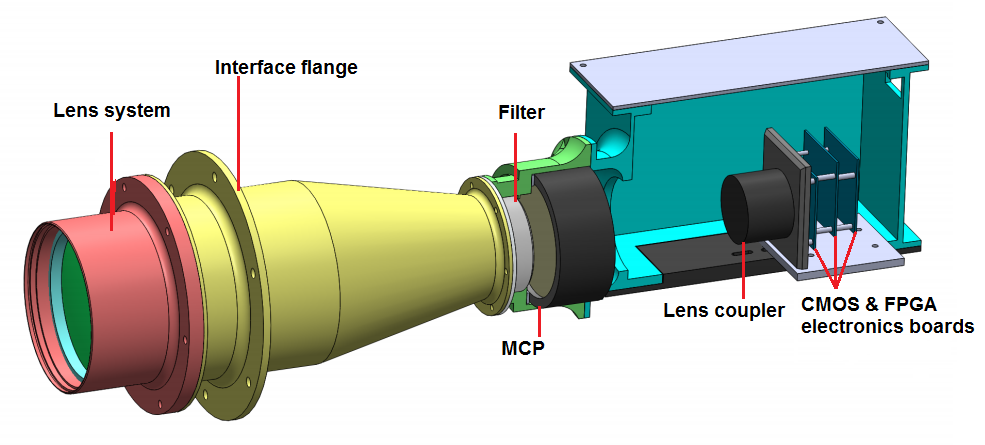}
\end{center}
\caption{Structural layout of the instrument}
\label{fig:Structural layout}
\end{figure*}
The lens barrel will be made up of Al-6061, because of its light weight and good structural properties. The mechanical structure basically consists of the lens system holder and the detector module. The structural layout is shown in Fig.~\ref{fig:Structural layout}. The optical cell consists of seven lenses (three single and two doublets) and spacers that position the lenses relative to each other. The payload interface of the instrument is at center of gravity (C.G.), so that it will give maximal stability.We designed the structure of the instrument to meet the most stringent requirements for launch vehicle platforms \citep{Ariane}. 

\begin{figure}[h!]
\begin{center}
\includegraphics[scale=0.06]{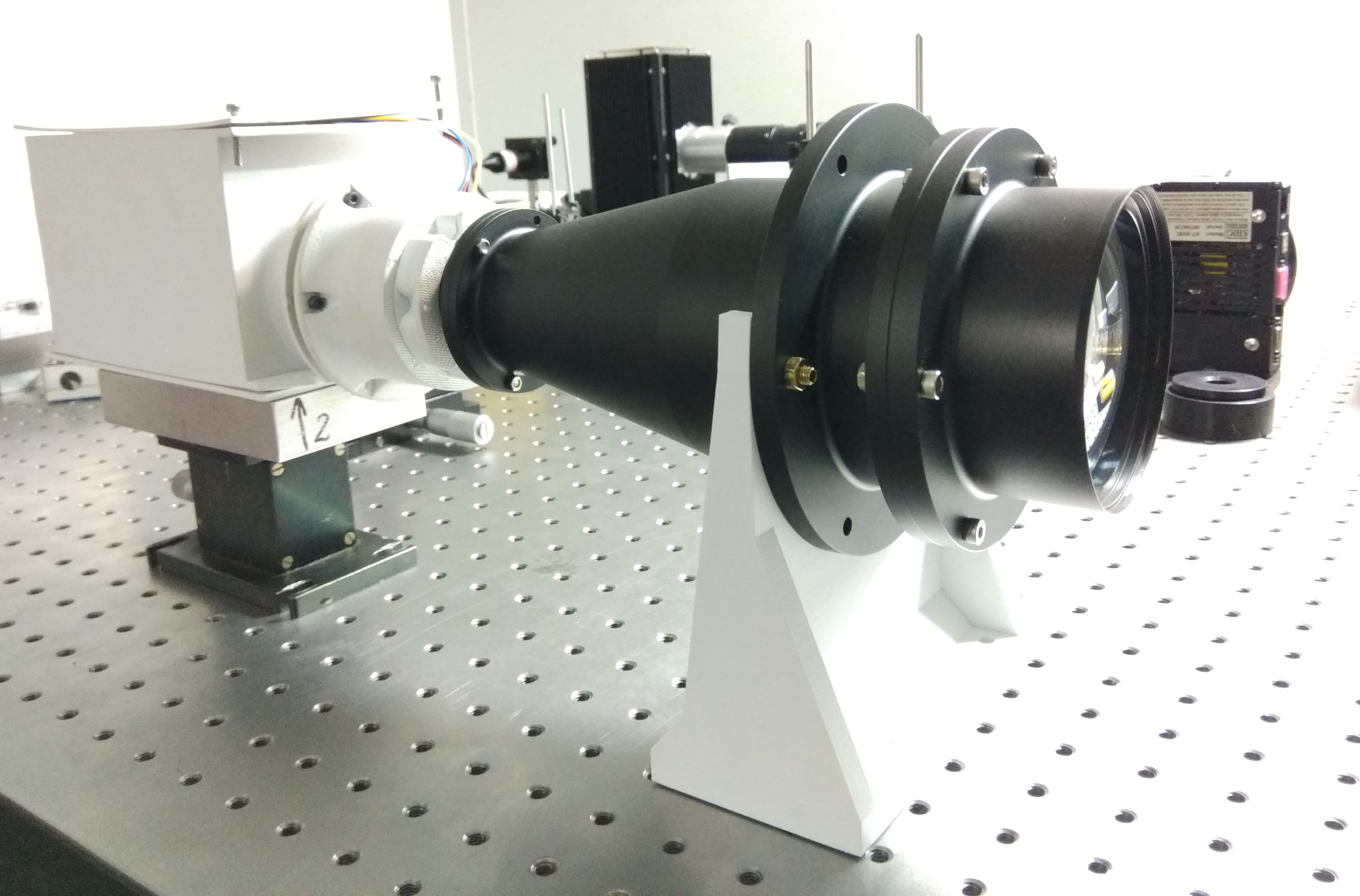}
\end{center}
\caption{Assembled instrument}
\label{fig:assembled_system}
\end{figure}

The final assembled instrument is shown in Fig.~\ref{fig:assembled_system}.

We have done a thermal analysis on the system, to see the optical performance of the instrument at various temperatures. The nominal operating temperature has to be between 18 to 22 degrees Celsius for best optical performance. 
To achieve the desired temperature requirements, an active thermal control system will be used, which consists of optical solar reflectors (OSR), heaters, and thermistors in a closed loop control system. For thermal isolation, the lens barrel will be covered with multi-layer insulation (MLI). The lens barrel is made up of aluminum, which is also black anodized will act as good radiator to radiate the heat through the baffle structure in front of the lens system. The high thermal conductivity of Aluminum will minimize the axial thermal gradient. To reduce the radial thermal gradients induced in the lens system, heaters will be spaced 120° around the lens barrel. In the CubeSat platform, we will be having a baffle with MLI layers and the front optics won't see direct radiation from the Sun, Moon or Earth albedo and this radiation will fall on the baffle surface. Since the heat flow due to radiation in the payload will be from the detector side to the baffle edge, the possible thermal gradient on the baffle, won't get transferred to the lens barrel side and in this way, the thermal gradient on the front optics would be taken care off. In the balloon platform, we will be using an additional dome structure with MLI blanket, around the rear lens to reduce the effects of thermal gradient.

\section{Flight Opportunities}

\subsection{High-Altitude Balloon platform}
We have developed the pointing and stabilization system for the balloon-born astronomical payloads \citep{Nirmal}, and the instrument will be mounted on this pointing platform for observations \citep{Sreejithballoon}.

\begin{figure}[h]
\begin{center}
\includegraphics[scale=0.55]{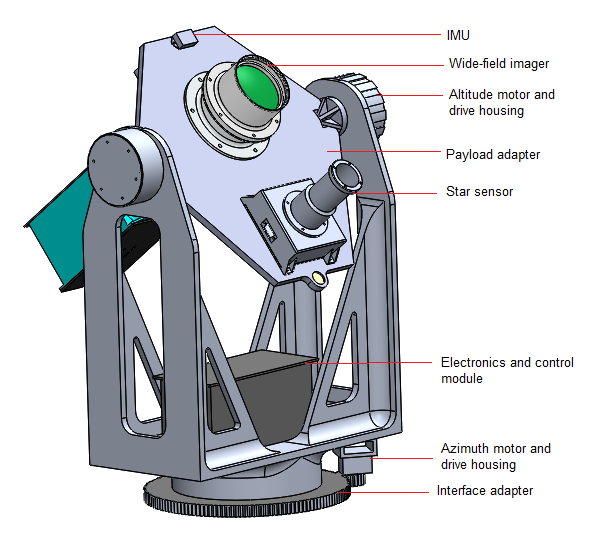}
\end{center}
\caption{Wide-field UV imaging payload on balloon-borne  pointing system}
\label{fig:Payload on balloon pointing system}
\end{figure}

The payload assembled on the pointing system is shown in Fig.~\ref{fig:Payload on balloon pointing system}. Pointing and stability of the system are carried out as a two-fold (coarse and fine pointing) operation, using inertial measurement sensors and a star sensor \citep{star_sensor}, respectively. Here, the attitude sensor obtains the pointing information and corrects the position to an accuracy of $\pm 0.24^{\circ}$ using servomotors, while the star sensor works in the inner loop, providing much finer accuracy of around $30^{\prime\prime}$. The attitude sensor acts as the brain for a closed-loop pointing system using servomotors. It combines an inertial measurement unit (consisting of 3-axis accelerometer, 3-axis gyroscope, and 3-axis magnetometer) with a GPS unit to give pointing to an accuracy of $\pm 0.24^{\circ}$ in either Earth-centered inertial coordinates (azimuth and elevation), or in the celestial equatorial coordinates (Right Ascension (RA) and Declination (Dec)). 

The fine pointing is achieved through the use of a star sensor, which is a highly sensitive wide-field imaging camera. The star sensor has a $10^{\circ}$ field of view with a limiting magnitude of $6.5$. It can provide pointing to an accuracy of $12.24^{\prime\prime}$ for S/N of $30$, and work at a maximum slew rate of about 2$^{\circ}$/sec. The coarse pointing maintains the payload at this slew rate, and the fine pointing is achieved by the inputs from the star sensor. The flight plan for this payload is on a zero-pressure balloon capable of floating at $\sim40$ km altitude for up to 5 hours with the National Balloon Facility station of Tata Institute of Fundamental Research Balloon Facility (TIFR), Hyderabad, later this year. The duration of the float will enable us to observe in the night a chosen astronomical object.

\subsection {Cubesat-based LEO orbit mission}

A CubeSat is a type of miniaturized satellite for space research that is made up of multiples of $10\times 10\times10$ cm cubic units (1U). CubeSats have mass of no more than 1.33 kilograms per unit, and often use commercial off-the-shelf (COTS) components for their electronics and structure. 
\begin{figure}[h!]
\begin{center}
\includegraphics[scale=0.23]{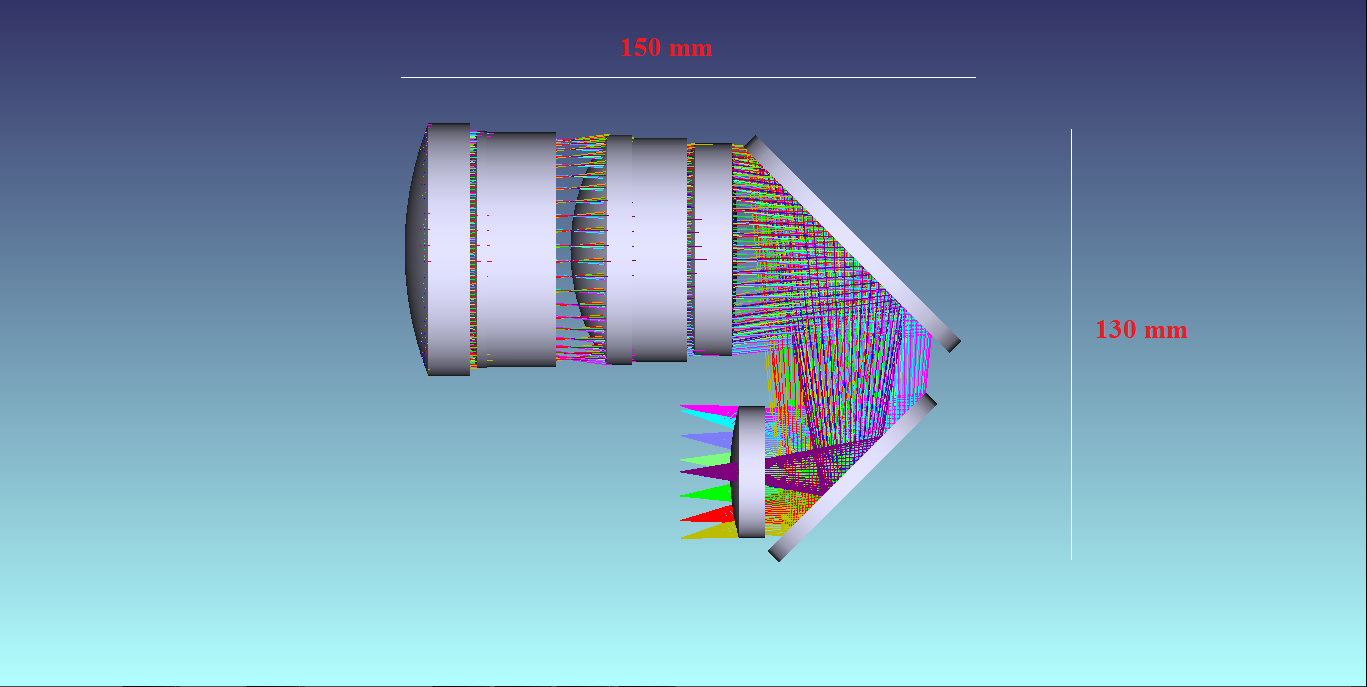}
\end{center}
\caption{Optical layout of UV imager to fit in 6U platform}
\label{fig:Optical layout of UV imager to fit in 6U platform}
\end{figure}

A CubeSat is an ideal platform for small aperture wide-field compact imaging experiment \citep{Broschsmall}. These days see affordable  launch opportunities for CubeSat-based payloads, and it is a cost-effective solution for space astronomy. We have designed the UV imager such that it can be also fit on a CubeSat platform.

\begin{figure}[h!]
\begin{center}
\includegraphics[scale=0.4]{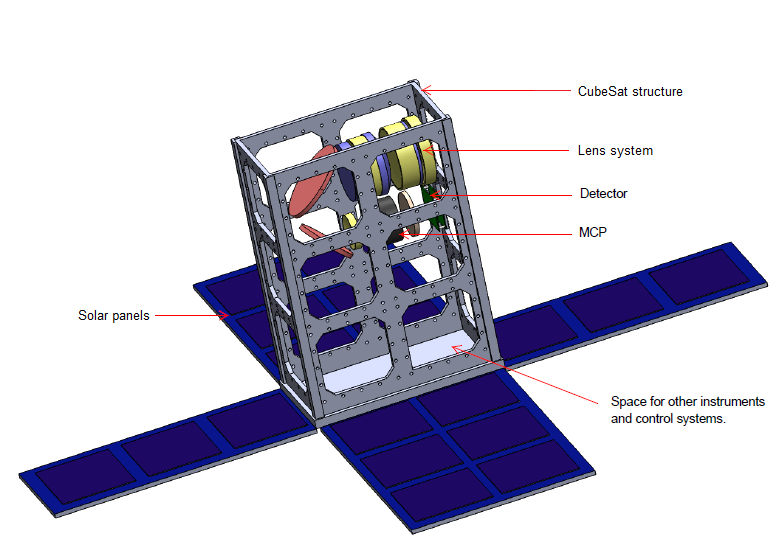}
\end{center}
\caption{UV imager on a 6U CubeSat platform}
\label{fig:Cubesat}
\end{figure}

After introducing two beam-folding flat mirrors (though the additional mirrors will reduce the optics efficiency by $\sim80$\% of the original configuration, 
provided 90\% of reflectivity for each flat mirror), the imaging system can fit on a 6U ($10\times20\times30$ cm) CubeSat (Fig.~\ref{fig:Optical layout of UV imager to fit in 6U platform}) platform. All the necessary power, telemetry and attitude control will be provided by the CubeSat platform. Fig.~\ref{fig:Cubesat} shows the UV imager on a 6U CubeSat.

\section{Conclusion}

Our instrument is a UV transient imager that combines a wide FOV with second-scale cadence and a possibility to stare at the same area of the sky for days, obtaining many epochs of observations.

\begin{table*}[ht!]
\begin{center}
\caption{Comparison of missions parameters in NUV channel.}
\label{table:parameters}
\resizebox{\textwidth}{!}{\begin{tabular}{lccccc}
\hline
  & \multicolumn{3}{c}{Past and Existing} & \multicolumn{2}{c}{Proposed} \\
  \midrule 
Mission parameter & GALEX & UVIT & WiFI & ULTRASAT & CUTIE \\\hline
Aperture (mm) & 500   &  376   & 80    &  330  &  80  \\
A$_{\rm coll}$ (cm$^2$) & 1950  & 851.78  & 50.3  & 856 &    50.3  \\ 
No. of pixels &4K$\times 4$K & $512\times 512$  &2K$\times 2$K  & 4K$\times 4$K &  40 Mpxl\\  
Pixel scale($^{\as}/$pix)  &1.5 &  3 &$7.5$ & 8.5 &  19 \\ 
Resolution ($^{\as}$) &  5.3  & 1.2--1.6 & $22$ &  $25$ &   $38$\\
FOV (deg$^2$)     & 1.21 & 0.18 &  91.6 & 235 &  121     \\
$\Delta\Omega$ (sr)  &$3.74\times 10^{-4}$&$5.21\times 10^{-5}$& $2.8\times 10^{-2}$ & $7.2\times 10^{-2}$ &$3.7\times 10^{-2}$\\ 
Range (nm) & $177-283$ &$200-300$ & $240-390$ & $220-280$  &$260-320$ \\   
Bandpass (nm)&  73.2  &  77 & 150  & 60 &   60  \\
$S$ (1/cm$^3$)             &$1.87\times 10^7$&$3\times 10^8$ & $6.2\times 10^6$ & $2.7\times 10^5$ & $8.9\times 10^6$     \\
Weight (kg)&    280   &    230 &  4   & 53.6  & $\sim 7$, for 6U CubeSat  \\ 
Cost  & \$150.6M & $\sim$\$6M & $\sim\$30,000$ &$> \$5$M & total cost a few \$M \\
Current status & Decommissioned & Active & Ready to fly &  Proposal reconsidered 2017& Postponed funding  \\
  &  2013     & Launched 2015              &        &               Launch 2023? & from 2015\\
\bottomrule
\end{tabular}}
\end{center}
\footnotesize{GALEX data from Morrisey et al. 2007. UVIT collecting area was calculated using available optical parameters. The data for range and bandpass are for the wideband filter. ULTRASAT data from Eran Ofek, ``Boutiques \& Experiments" meeting, Caltech, 2016. CUTIE data source: NASA TechPort Data Sheet (2014-2015) and Cenko et al. (2017).}
\end{table*}

Such a mission has the potential to make both expected and unexpected discoveries in the NUV transient domain at short timescales. In Table~\ref{table:parameters} we compare the general mission parameters for different space missions with NUV capabilities, existing and proposed. 

In addition, we have calculated the survey parameter for comparing the ability of UV missions to estimate the UV background \citep{henry1982,surveypar},
\be
S=\fr{100}{A\D\Omega \D\lambda}\,,
\label{eq:surveypar}
\ee
where $A$ is the collecting aperture in cm$^2$, $\D\Omega$ is the FOV in sr, and $\D\lambda$ is the bandpass.

We can confidently say that our instrument is comparable to many proposed wide-field UV imagers in detecting transient events. For example, M dwarfs account for more than 75\% of the stellar population in the solar neighbourhood (up to 1 Kpc). A vast majority of UV short transients is associated
with stellar flare eruptions on K and M stars. 
Such a flare on, for example, Proxima Cen, could correspond to a
flux at Earth of $10^5$--$10^6$ UV phots/cm$^2$ for a period
of several hours. The M dwarf flare 100 pc away will
produce UV flux at Earth of $10$--$10^2$ phots/cm$^2$, while our detection limit is $1.34\times 10^{-4}$ photons/cm$^2$/sec (see Table~\ref{table:fluxes}). Number of detected flares can be estimated using
\be
N_{\rm F} = \fr{R_{\rm F}\tau\Delta\Omega}{4\pi}\,,
\label{eq:flares}
\ee
where $R_{\rm F}$ is the annual rate of flares and $\tau$ exposure per survey/FOV. For M dwarfs flares, the all-sky rate was estimated to be $R_{\rm F}^{\rm dM}=10^8/$yr (Kulkarni \& Rau 2006). Rate of collisions in planetary systems were estimated in Safonova, Sivaram \& Murthy (2008) as $R_{\rm F}^{\rm coll} \simeq 10^9/$yr. 

This gives the rates as 0.7 and 7, respectively,  in the WiFI FOV in 100 sec planned stare/survey exposure time. As for the tidal disruption flares (TDFs), different sources estimate different rates, starting with $1\times 10^{-4}$/yr (Gezari et al. 2008), to 1/yr/20 deg$^2$ (Gezari et al. 2012) based on a TDS sensitivity. With the same sensitivity as TDS survey (10 AB mag) and FOV of 91.6 deg$^2$, we can get the detection rate of 4.6/yr per FOV and a planned 100 sec survey exposure. However, with our sensitivity at 18 AB mag, we can improve this rate by order of magnitude. Another interesting type of flare that we have discussed already in Safonova et al. (2008) -- stellar-planetary collisions -- has recently been given impetus with the discovery of a binary system HD240430 and HD240429, where one of the stars is metal-enriched. It can be explained by the engulfment of an equivalent of 15 Earths (Oh et al. 2017). 
\begin{table}[h!]
\begin{center}
\caption{Approximate NUV fluxes at Earth from different flares.}
\label{table:fluxes}
\small
\begin{tabular}{|c|c|c|}
\toprule
NUV flux from (ph/cm$^2$/s)  & at 10 pc    & at 1 Kpc   \\
\hline
Host star (G, K) & $1-50$  & $10^{-2}-10^{-3}$  \\
Planet-planet coll. event & upto $10^3$   & upto $1$  \\
\hline
stellar-planetary collisions & at 10 pc & at 10 Kpc \\
   & $\sim 10^8$ & $\sim 10^2$ \\
\hline
dM flare  &  at 1 pc  & at 100 pc\\
   & $10^5-10^6$ & $10-10^2$  \\
\hline
\end{tabular}\end{center} 
\end{table}

We have presented the design of a wide-field UV imager to study the transient events in the NUV domain. The instrument has a wide FOV of 10.8 degrees with $22^{\prime\prime}$ resolution. The opto-mechanical design has taken into accounting the primary limiting conditions such as, volume, weight constraints, launch load vibrations, compactness and cost effectiveness. The imager is designed in such a way that its lightweight and compact nature fits well a CubeSat configuration and we are considering a possibility of a LEO flights on CubeSats. We are planning to replace the current detector with a solar blind CsTe photocathode-based detector to avoid any red leak. In addition, the telescope can be flown on any nanosatellite as a piggyback payload, where one of the main scientific objectives will be the study of the transient UV sky. As a qualification  flight, we will fly this instrument on-board the high-altitude balloon to perform observations in the NUV domain from the floating altitude of about 40 km.

\section {Acknowledgements}

Part of this research has been supported by the Department of Science and Technology (Government of India) under Grant IR/S2/PU-006/2012. A.~G.~Sreejith acknowledges Austrian Forschungsförderungsgesellschaft FFG project “ACUTEDIRNDL” P859718.


\end{document}